\documentstyle[12pt]{article}

\newcommand{\be}{\begin{equation}}
\newcommand{\ee}{\end{equation}}

\begin{document}
\begin{center}
{\it International Journal of Modern Physics B 19 (2005) 879-897}
\vskip 7mm
{\Large \bf Thermodynamics of Few-Particle Systems }

\vskip 7mm
{\large \bf Vasily E. Tarasov }

{\it Skobeltsyn Institute of Nuclear Physics, \\
Moscow State University, Moscow 119992, Russia}

E-mail: tarasov@theory.sinp.msu.ru 
\end{center}

\begin{abstract}
We consider the wide class of few-particle systems 
that have some analog of the thermodynamic laws. 
These systems are characterized by the distributions 
that are determined by the Hamiltonian and satisfy the 
Liouville equation. 
Few-particle systems of this class are described by
a non-holonomic constraint: the power of non-potential forces
is directly proportional to the velocity of 
the elementary phase volume change.  
The coefficient of this proportionality is determined by the Hamiltonian. 
In the general case, the examples of the few-particle systems of 
this class are the constant temperature systems, 
canonical-dissipative systems, and Fermi-Bose 
classical systems. 
\end{abstract}

\vskip 3 mm

\section{Introduction}

The main aim of statistical thermodynamics is to derive the 
thermodynamic properties of systems starting from a description 
of the motion of the particles. 
Statistical thermodynamics of few-particle systems have recently 
been employed to study a wide variety of problems
in the field of molecular dynamics. 
In molecular dynamics calculations, few-particle systems 
can be exploited to generate statistical ensembles as 
the canonical, isothermal-isobaric and isokinetic ensembles 
\cite{E,EHFML,HG,EM,Nose1,Nose2,Nose,Tuck2}.  
The aim of this work is the extension of the statistical
thermodynamics to a wide class of few-particle systems.
We can point out some few-particle systems  
that have some analog of thermodynamics laws.

\begin{itemize}
\item The constant temperature systems with minimal Gaussian constraint  
are considered in Ref. \cite{E,EHFML,HG,EM,Nose}.
These systems are the few-particle systems
that are defined by the non-potential forces in the form
${\bf F}^{(n)}_i=-\gamma {\bf p}_i$ and the 
Gaussian non-holonomic constraint. 
This constraint can be represented 
as an addition term to the non-potential force.

\item The canonical distribution can be derived 
as a stationary solution of the Liouville equation 
for a wide  class of few-particle system \cite{mplb}.
This class is defined by a very simple condition 
for the non-potential forces:
the power of the non-potential forces must be directly proportional
to the velocity of the Gibbs phase (elementary phase volume) change.
This condition defines the general constant temperature systems.
This condition leads to the canonical distribution 
as a stationary solution of the Liouville equations.
For the linear friction, we derived the constant temperature systems. 
The general form of the non-potential forces 
is not considered in Ref. \cite{mplb}.

\item The canonical-dissipative systems are described in Ref. \cite{Eb,SET}.
These systems are the few-particle systems 
are defined by the non-potential forces 
${\bf F}^{(n)}_i=-\partial G(H)/ \partial {\bf p}_i$,
where $G(H)$ is a function of Hamiltonian $H$. 
The distribution functions are derived as 
solutions of the Fokker-Planck equation. 
Note that Fokker-Planck equation can be derived
from the Liouville equation \cite{Is}.

\item The quantum few-particle systems
with pure stationary states are suggested in Ref. \cite{Tarpre02,Tarpla02}. 
The correspondent classical systems are not discussed.

\item The few-particle systems with the fractional
phase space and non-Gaussian distributions are suggested in \cite{chaos,PRE05}. 
Note that  nondissipative systems with the usual phase space
are dissipative systems in the fractional phase space \cite{chaos,PRE05}.
\end{itemize}

The analog of the first law of thermodynamics is connected 
with the variation of the mean value Hamiltonian
\be U(x)=\int H({\bf q},{\bf p},x) \rho({\bf q}, {\bf p},x) 
d^N{\bf q} d^N{\bf p}  \ee
that has the form
\[ dU(x)=\int \delta_{x} H \rho d^N{\bf q} d^N{\bf p}
+\int H \delta_{x}\rho  d^N{\bf q} d^N{\bf p}. \]
We can have the analog of the second laws of thermodynamics
if the second term on the right-hand side  can be represented in the form
\[ T(x)dS(x)=T(x) \delta_x \int S_N (\rho) \rho d^N{\bf q} d^N{\bf p} . \]
In this case, we have the condition
\[ H \delta_{x}\rho = T(x) \delta_x (\rho S_N(\rho)).  \]
This representation is realized if the distribution $\rho$ 
can be represented as a function of Hamiltonian $\rho=\rho(H,x)$  
such that we can write $H=T(x) G(\rho)$. 
Obviously, we have the second requirement for this distribution:
the distribution $\rho$ satisfy the Liouville 
equation of the systems. For these N-particle systems, we can use
the analogs of the usual thermodynamics laws.  
Note that $N$ is an arbitrary natural number since we do not
use the condition $N \gg 1$ or $N \rightarrow \infty$.
This allows us to use the suggested few-particle systems 
for the simulation schemes \cite{FS} 
for the molecular dynamics.

In this paper we consider few-particle systems with distributions 
that are defined by Hamiltonian and Liouville equation. 
We describe the few-particle systems that have some analog of 
the thermodynamic laws. 
These systems can be defined by the non-holonomic (non-integrable)
constraint: the power of non-potential forces is 
directly proportional to the velocity of the elementary phase volume change.  
In the general case, the coefficient of this proportionality is 
determined by the Hamiltonian. 
The special constraint allows us to derive distributions for the system, 
even in far-from equilibrium states. 
The examples of these few-particle systems are
the constant temperature systems \cite{E,EHFML,HG,EM,Nose1,Nose2,Nose,Tuck2}, 
the canonical-dissipative systems \cite{Eb,SET}, and the Fermi-Bose 
classical systems \cite{Eb}.


In Sec. 2, we derive the analog of thermodynamic laws 
for the few-particle systems with the distributions that are
defined by Hamiltonian.
In Sec. 3, we consider  
the condition for the non-potential forces
that allows us to use the analog of thermodynamic laws. 
We consider the wide class of few-particle systems with canonical 
Gibbs distribution and non-Gaussian distributions.
In Sec. 4, we consider the non-holonomic constraint for
few-particle systems. 
We formulate the proposition which allows us to derive 
the thermodynamic few-particle systems from the equations 
of few-particle system motion.
The few-particle systems with the simple Hamiltonian 
and the simple non-potential forces are considered. 
Finally, a short conclusion is given in Sec. 5.

\section{Thermodynamics Laws}

\subsection{First Thermodynamics Law}

Let us consider the N-particle classical system in the Hamilton picture 
and denote the position of the $i$th particle 
by ${\bf q}_i$ and its momentum by ${\bf p}_i$, where $i=1,...,N$. 
The state of this system is described by the distribution function 
$\rho=\rho({\bf q},{\bf p},x,t)$. 

The mean value of the function
$f({\bf q},{\bf p},x,t)$ is defined by the following equation
\be f(x,t)=
\int f ({\bf q},{\bf p},x,t) \rho({\bf q}, {\bf p},x,t) 
d^N{\bf q} d^N{\bf p} . 
\ee
Here, $x=\{x_1,x_2,...x_n\}$ are external parameters.
The variation $\delta_x$ for this function can be defined by the relation
\be \label{var-x}
\delta_x f ({\bf q},{\bf p},x,t)=\sum^{n}_{k=1} 
\frac{\partial f ({\bf q},{\bf p},x,t)}{\partial x_k} dx_k .
\ee

The first law of thermodynamics states that the internal energy $U(x)$
may change because of (1) heat transfer $\delta Q$, and (2) work $\delta A$
of thermodynamics forces $X_k(x)$:
\be \delta A=\sum^n_{k=1} X_k(x) d x_k . \ee
The external parameters $x$ here act as generalized coordinates.
In the usual equilibrium thermodynamics the work done does 
not entirely account for the change in the internal energy $U(x)$. 
The internal energy also changes because of the transfer of heat, 
and so
\be dU=\delta Q- \delta A . \ee
Since thermodynamic forces $X_k(x)$ are non-potential forces
\be \label{FaFa}
\frac{\partial X_k (x)}{\partial x_l}=
\frac{\partial X_l (x)}{\partial x_k} , 
\ee
the amount of work $\delta A$ depends on the path of transition 
from one state in parameters space to another.
For this reason $\delta A$ and $\delta Q$, taken separately, 
are not total differentials.

Let us give statistical definitions of internal energy, thermodynamic 
forces and heat transfer for the few-particle systems in the mathematical 
expression of the analog of the first thermodynamics law. 
It would be natural to define the internal energy
as the mean value of Hamiltonian $H$:
\be 
U(x)=\int H({\bf q},{\bf p},x) \rho({\bf q}, {\bf p},x) 
d^N{\bf q} d^N{\bf p} . \ee
It follows that the expression for the total differential
has the form
\[ dU(x)=\int \delta_{x} H({\bf q},{\bf p},x) 
\rho({\bf q}, {\bf p},x) d^N{\bf q} d^N{\bf p}
+\int H({\bf q},{\bf p},x) 
\delta_{x}\rho({\bf q}, {\bf p},x) d^N{\bf q} d^N{\bf p}. \]
Using Eq. (\ref{var-x}), we have 
\be \label{dU} 
dU(x)=\int \frac{\partial H({\bf q},{\bf p},x)}{\partial x_k} 
\delta x_k \rho({\bf q}, {\bf p},x) d^N{\bf q} d^N{\bf p}
+\int H({\bf q},{\bf p},x) \delta_{x}\rho({\bf q}, {\bf p},x) 
d^N{\bf q} d^N{\bf p} . \ee

In the first term on the right-hand side we can use the 
definition of phase density of the thermodynamic force
\[  X^{d}_{k}({\bf q},{\bf p},x)=-
\frac{\partial H({\bf q},{\bf p},x)}{\partial x_k} . \]
The thermodynamic force $X_k(x)$ is the mean value of 
the phase density of the thermodynamics force
\be \label{Fia}
X_{k}(x)=\int 
X^{d}_{k}({\bf q},{\bf p},x)\rho({\bf q}, {\bf p}, x) 
d^N{\bf q} d^N{\bf p} .
\ee
Using this equation, we can prove relation (\ref{FaFa}).

Analyzing these expressions,  we see that the first term on the 
right-hand side of differential (\ref{dU}) answers for 
the work of thermodynamics forces (\ref{Fia}), 
whereas the amount of the heat transfer is given by
\be \label{dQ} \delta Q=
\int H({\bf q},{\bf p},x) \delta_{x}\rho({\bf q}, {\bf p}, x) 
d^N{\bf q} d^N{\bf p} . \ee
We see that the heat transfer term accounts for the change in the
internal energy due not to the work of thermodynamics forces, but
rather to the change in the distribution function
cased by the external parameters $x$.

\subsection{Second Thermodynamics Law}

Now let us turn our attention to the analog of the second law 
for the few-particle systems. 

The second law of thermodynamics has the form
\be \label{SL}
\delta Q=T(x) dS(x) . \ee
This implies that there exists a function of state $S(x)$
called entropy.
The function $T(x)$ acts as an integration factor.
Let us prove that the law (\ref{SL}) follows from the statistical 
definition of $\delta Q$ in Eq. (\ref{dQ}).
For Eq. (\ref{dQ}), we take the distribution that is defined by 
the Hamiltonian, and show that Eq. (\ref{dQ}) 
can be reduced to Eq. (\ref{SL}).

We can have the analog of the second laws of thermodynamics if 
the right-hand side of Eq. (\ref{dQ}) can be represented in the form
\[ T(x)dS(x)=T(x) \delta_x \int \rho({\bf q}, {\bf p},x) 
S_N(\rho({\bf q},{\bf p},x))  d^N{\bf q} d^N{\bf p} . \]
This requirement can be written in an equivalent form
\[ H \delta_{x}\rho = T(x) \delta_x (\rho S_N(\rho)).  \]
This representation is realized if the distribution $\rho$ 
can be represented as a function of Hamiltonian 
\be \label{rhoH0} \rho({\bf q},{\bf p},x)=\rho(H({\bf q},{\bf p},x),x) .\ee
Obviously, we have the second requirement for the distribution $\rho$,
which must satisfies the Liouville equation.
We assume that Eq. (\ref{rhoH0}) can be solved  in the form 
\[ H=T(x) G(\rho) , \]
where $G$ depends on the distribution $\rho$. The function
$T(x)$ is a function of the parameters $x=\{x_1,x_2,...,x_n\}$. 
As a result, we can rewrite Eq. (\ref{dQ}) in the equivalent form
\be \label{dQ2} \delta Q= 
\int \Bigl( T(x) G(\rho({\bf q},{\bf p},x),x)
+H_0(x) \Bigr)
\delta_{x}\rho({\bf q}, {\bf p},x) d^N{\bf q} d^N{\bf p} .
\ee
The term with $H_0(x)$, which is added into this equation, is equal to
zero because of the normalization condition
of distribution function $\rho$:
\[ H_0(x) \delta_{x} \int \rho({\bf q}, {\bf p},x) d^N{\bf q} d^N{\bf p}=
H_0(x) \delta_{x} 1=0 . \]
For canonical Gibbs distribution
\[ \rho=\frac{1}{Z(x)} \exp -\frac{H({\bf q},{\bf p},x)}{kT(x)} , \]
where $Z(x)$ is defined by
\[ Z(x)=\int \exp -\frac{H({\bf q},{\bf p},x)}{kT(x)} \ d^N {\bf q} d^N {\bf p}, \]
we use $G(\rho)$ and $H_0(x)$ in the form
\[ H_0(x)=kT(x) ln Z(x)-kT(x), \quad G(\rho)=-k ln (Z(x) \rho) . \]
As a result, Eq. (\ref{dQ2}) can be rewritten in the form
\be \label{dQ3}
\delta Q= T(x)\delta_{x}\int \rho({\bf q},{\bf p},x) 
S_N(\rho({\bf q},{\bf p},x)) d^N{\bf q} d^N{\bf p} , \ee
where the function $S_N(\rho)$ is defined by
\be
\frac{\partial (\rho S_N(\rho))}{\partial \rho}=G(\rho)+H_0(x)/T(x) .
\ee
We see that the expression for $\delta Q$ is integrable.
If we take $1/T(x)$ for the integration factor, thus 
identifying $T(x)$ with the analog of absolute temperature,
then, using Eqs. (\ref{SL}) and (\ref{dQ3}), we can give 
the statistical  definition of entropy: 
\be \label{SaT} S(x)=
\int \rho({\bf q},{\bf p},x) 
S_N(\rho({\bf q},{\bf p},x)) d^N{\bf q} d^N{\bf p} +S_0. \ee
Here $S_0$ is the contribution to the entropy which does not depend
on the variables $x$, but may depend on the number of particles $N$
in the system.
As a result, the expression for entropy is equivalent to
the mean value of phase density function
$S^{d}({\bf q},{\bf p},x)=S(\rho({\bf q},{\bf p},x) )+S_0$. 
Here $S^{d}$ is a function of dynamic variables 
${\bf q},{\bf p}$, and the parameters $x=\{x_1,x_2,...,x_n\}$.
The number  $N$ is an arbitrary natural number since we do not
use the condition $N\gg 1$ or $N \rightarrow \infty$.
Note that in the usual equilibrium thermodynamics 
the function $T(x)$ is a mean value of kinetic energy.
In the suggested thermodynamics of few-particle systems 
$T(x)$ is the usual function of the external parameters 
$x=\{x_1,x_2,...,x_n\}$.

\subsection{Thermodynamic Few-Particle Systems}

Let us define the special class of the few-particle systems with 
distribution functions that are completely characterized by the Hamiltonian. 
These distributions must satisfy 
the Liouville equation for the few-particle system. 

{\bf Definition}
{\it A few-particle system 
\[ \frac{d{\bf q}_i}{dt}=\frac{\partial H}{\partial {\bf p}_i},
\quad
\frac{d{\bf p}_i}{dt}=-\frac{\partial H}{\partial {\bf q}_i}+
{\bf F}^{(n)}_i  \]
will be called a thermodynamic few-particle system if the 
following conditions are satisfied: \\
(1) the distribution function   $\rho$
is determined by the Hamiltonian, 
i.e., $\rho(q,p,x)$ can be written in the form
\be \label{rhoH} \rho({\bf q},{\bf p},x)=\rho(H({\bf q},{\bf p},x),x), \ee
where $x$ is a set of external parameters, and 
\[ \rho ({\bf q},{\bf p},x) \ge0, \quad 
\int \rho({\bf q},{\bf p},x) d^N {\bf q} d^N {\bf p}=1 ;\]
(2) the distribution function   $\rho$ 
satisfies the Liouville equation 
\[ \frac{\partial \rho}{\partial t}+
\frac{\partial}{\partial {\bf q}_i}\Bigl(\frac{\partial H}{\partial {\bf p}_i}
 \rho \Bigr)+
\frac{\partial}{\partial {\bf p}_i} \Bigl((-\frac{\partial H}{\partial {\bf q}_i}+
{\bf F}^{(n)}_i ) \rho \Bigr)=0 ; \]
(3) the number of particles $N$ is an arbitrary natural number. }\\

Here and later we mean the sum on the repeated index $i$ from 1 to N.

Examples of the thermodynamic few-particle systems:\\

\noindent
(1) The constant temperature systems \cite{E,EHFML,HG,EM,Nose1,Nose2,Nose,Tuck2}
that have the canonical distribution. In general,
these systems can be defined by the non-holonomic 
constraint, which is suggested in Ref. \cite{mplb}. \\
(2) The classical system with the Breit-Wigner 
distribution function that is defined by
\be \label{BreitWigner}
\rho(H({\bf q},{\bf p},x))=\frac{\lambda}{(H({\bf q},{\bf p},x)-E)^2+(\Gamma/2)^2} .
\ee
(3) The classical Fermi-Bose canonical-dissipative systems \cite{Eb} 
that are defined by the distribution functions in the form
\be \label{FermiBose}
\rho(H({\bf q},{\bf p},x))=\frac{1}{\exp [\beta(x) (H({\bf q},{\bf p},x)-\mu)]+a} .
\ee

\section{Distribution for Thermodynamic Few-Particle Systems}

\subsection{Formulation of the Results}

Let us consider the few-particle systems which are defined by the equations
\be \label{Sys}
\frac{d{\bf q}_i}{dt}=\frac{\partial H}{\partial {\bf p}_i},
\quad
\frac{d{\bf p}_i}{dt}=-\frac{\partial H}{\partial {\bf q}_i}+
{\bf F}^{(n)}_i , \ee
where $i=1,...,N$. The power of non-potential forces ${\bf F}^{(n)}_i$ 
is defined by
\be \label{power} {\cal P}({\bf q},{\bf p},x)=
{\bf F}^{(n)}_i \frac{\partial H}{\partial {\bf p}_i}. \ee
If the power of the non-potential forces is equal to zero 
(${\cal P}=0$) and $\partial H/ \partial t=0$, then few-particle system
is called a conservative system. 
The velocity of an elementary phase volume change $\Omega$ is defined 
by the equation
\be \label{omega} \Omega({\bf q},{\bf p},x)=
\frac{\partial {\bf F}_i}{\partial {\bf p}_i}+
\frac{\partial^2 H}{\partial {\bf q}_i \partial {\bf p}_i}=
\frac{\partial {\bf F}^{(n)}_i}{\partial {\bf p}_i} . \ee
We use the following notations for the scalar product
\[ \frac{\partial {\bf A}_{i}}{\partial {\bf a}_i}=
\sum^N_{i=1}\Bigl(
\frac{\partial A_{xi}}{\partial a_{xi}}+
\frac{\partial A_{yi}}{\partial a_{yi}}+
\frac{\partial A_{zi}}{\partial a_{zi}}\Bigr). \]

The aim of this section is to prove the following result. \\

{\bf Proposition 1.}
{\it If the non-potential forces ${\bf F}^{(n)}_i$ of 
the few-particle system  (\ref{Sys}) 
satisfy the constraint condition
\be \label{NC-P1} 
\frac{\partial {\bf F}^{(n)}_i}{\partial {\bf p}_i}-
\beta(H,x) {\bf F}^{(n)}_i \frac{\partial H}{\partial {\bf p}_i}
=0, \ee
then this system is a thermodynamic few-particle system 
with the distribution function
\be \label{distr}
\rho({\bf q},{\bf p},x)=\frac{1}{Z(x)} \exp - B(H({\bf q},{\bf p},x),x) ,
\ee
where the function $B(H,x)$ is defined by the equation
$\partial B(H,x)/ \partial H=\beta(H,x)$, and }  \\
\[ Z(x)=\int \exp[ - B(H({\bf q},{\bf p},x),x)] d^N {\bf q} d^N {\bf p}\]

Obviously, we consider the distribution functions (\ref{distr}) 
and the function $B=B(H,x)$ such that $\rho \ge0$, and $Z(x) < \infty$.
Note that condition (\ref{NC-P1}) means that 
the velocity of the elementary phase volume change 
$\Omega$ is directly proportional to the power ${\cal P}$ 
of non-potential forces ${\bf F}^{(n)}_i$ of the few-particle 
system (\ref{Sys}) and coefficient of this proportionality
is a function $\beta(H,x)$ of Hamiltonian $H$, i.e., 
\be \label{NC-P0} \Omega({\bf q} ,{\bf p} ,x)-
\beta(H,x) {\cal P}({\bf q} ,{\bf p} ,x)=0 . \ee

Note that any few-particle system with the non-holonomic constraint 
(\ref{NC-P0}) or (\ref{NC-P1}) is a thermodynamic few-particle system. 
Solving the Liouville equation with the non-holonomic constraint 
(\ref{NC-P1}), we can obtain the 
distributions that are defined by the Hamiltonian.

\subsection{Proof of Proposition 1}

Let us consider the Liouville equation for the few-particle
distribution function $\rho=\rho({\bf q},{\bf p},x,t)$. 
This distribution function $\rho({\bf q},{\bf p},x,t)$ 
express a probability that a phase space point $({\bf q},{\bf p})$ 
will appear. The Liouville equation for this few-particle system
\be \label{rhoN} \frac{\partial \rho}{\partial t}+
\frac{\partial}{\partial {\bf q}_i}\Bigl({\bf K}_i \rho \Bigr)+
\frac{\partial}{\partial {\bf p}_i} \Bigl({\bf F}_i \rho \Bigr)=0\ee
expresses the conservation of probability in the phase space. 
Here we use
\[ {\bf K}_i=\frac{\partial H}{\partial {\bf p}_i} , \quad
{\bf F}_i=-\frac{\partial H}{\partial {\bf q}_i}
+{\bf F}^{(n)}_i .\]
Using a total time derivative along the phase space trajectory by
\be \frac{d}{dt}=
\frac{\partial}{\partial t}+
{\bf K}_i \frac{\partial}{\partial {\bf q}_i}+
{\bf F}_i \frac{\partial}{\partial {\bf p}_i} , \ee
we can rewrite Eq. (\ref{rhoN}) in the form:
\be \label{39} \frac{d\rho}{dt}=-\Omega \rho , \ee
where the omega function is defined by Eq. (\ref{omega}). 
In classical mechanics of Hamiltonian systems the right-hand side of 
the Liouville equation (\ref{39}) is zero, and 
the distribution function does not change with time.  
For the N-particle systems (\ref{Sys}), 
the omega function (\ref{omega}) does not vanish.
For this system, the omega function is defined by (\ref{omega}). 
For the thermodynamic few-particle systems, this function 
is defined by the constraint (\ref{NC-P1}) in the form
\be \label{A1} 
\Omega=\beta(H,x) {\bf F}^{(n)}_i \frac{\partial H}{\partial {\bf p}_i}. 
\ee
In this case, the Liouville equation has the form
\be \label{31}
\frac{d\rho({\bf q},{\bf p},x)}{dt}=-
\beta(H,x) {\bf F}^{(n)}_i \frac{\partial H}{\partial {\bf p}_i} \rho.
\ee
Let us consider the total time derivative of the Hamiltonian.  
Using equations of motion (\ref{Sys}), we have 
\be \label{A2} 
\frac{dH}{dt}=\frac{\partial H}{\partial t}+
\frac{\partial H}{\partial {\bf p}_i}
\frac{\partial H}{\partial {\bf q}_i}+
\Bigl(-\frac{\partial H}{\partial {\bf q}_i}
+{\bf F}^{(n)}_i \Bigr)\frac{\partial H}{\partial {\bf p}_i} = 
 \frac{\partial H}{\partial t}+
{\bf F}^{(n)}_i \frac{\partial H}{\partial {\bf p}_i} . \ee
If ${\partial H}/{\partial t}=0$, then 
the power ${\cal P}$ of non-potential forces is equal to 
the total time derivative of the Hamiltonian
\[ {\bf F}^{(n)}_i \frac{\partial H}{\partial {\bf p}_i}=
\frac{dH}{dt} . \]
Therefore Eq. (\ref{31}) can be written in the form
\be \label{Eq45} 
\frac{d ln \ \rho({\bf q},{\bf p},x)}{dt}=
-\beta(H,x) \frac{dH}{dt}. \ee
If $\beta(H,x)$ is an integrable function, then this function 
can be represented as a derivative 
\be  \label{L} \beta(H,x)=\frac{\partial B(H,x)}{\partial H} . \ee
In this case, we can write Eq. (\ref{Eq45}) in the form
\be  \frac{d ln \ \rho({\bf q},{\bf p},x)}{dt}=- \frac{dB(H,x)}{dt} . \ee
As a result, we have the following solution of the Liouville equation 
\be \label{solution}
\rho({\bf q},{\bf p},x)=\frac{1}{Z(x)} \exp - B(H({\bf q},{\bf p},x),x) .
\ee
The function $Z(x)$ is defined by the normalization condition. 
It is easy to see that the distribution function of the 
N-particle system is determined by the Hamiltonian.
Therefore, this system is a thermodynamic few-particle system.
Note that $N$ is an arbitrary natural number since we do not
use the condition $N\gg 1$ or $N \rightarrow \infty$.

\subsection{Few-Particle Systems with Canonical Distributions}

In this section, we consider the thermodynamic few-particle system 
that is described by canonical distribution \cite{mplb}.  
These few-particle systems are defined by 
the simple function $\beta(H,x)=3N \beta(x)$ 
in the non-holonomic constraint (\ref{NC-P0}). \\

{\bf Corollary 1.}
{\it If velocity of the elementary phase volume change $\Omega$
is directly proportional to the power of non-potential forces ${\cal P}$, 
then we have the usual canonical Gibbs distribution as a 
solution of the Liouville equation. } \\

In other words, the few-particle system with the
non-holonomic constraint $\Omega=\beta(x) {\cal P}$
can have the canonical Gibbs distribution 
\[ \rho({\bf q},{\bf p},x)=\exp \, \beta(x) \Bigl(
{\cal F}(x)- H({\bf q} ,{\bf p} ,x) \Bigr) \]
as a solution of the Liouville equation.
Here the coefficient $\beta(x)$ does not depend
on $({\bf q} ,{\bf p} ,t)$, i.e., $d\beta(x)/dt=0$.
Proof of this corollary is considered in Ref. \cite{mplb}.

Using Eq. (\ref{power}), we get
the Liouville equation in the form
\be \label{91}
\frac{d\rho({\bf q},{\bf p},x)}{dt}=-
\beta(x) {\bf F}^{(n)}_i \frac{\partial H}{\partial {\bf p}_i} \rho.
\ee
The total time derivative for the Hamiltonian is defined by Eq. (\ref{A2})
in the form
\[ \frac{dH}{dt}=\frac{\partial H}{\partial t}+
{\bf F}^{(n)}_i \frac{\partial H}{\partial {\bf p}_i} . \]
If ${\partial H}/{\partial t}=0$, then the energy change 
is equal to the power ${\cal P}$ of the
non-potential forces ${\bf F}^{(n)}_i$.  
Therefore the Liouville equation can be rewritten in the form
\[ \frac{d \ ln \rho({\bf q} ,{\bf p} ,x)}{dt}+
\beta(x)\frac{dH({\bf q} ,{\bf p} ,x)}{dt}=0. \]
Since coefficient $\beta(x)$ is a constant ($d\beta(x)/dt=0$), we have
\[ \frac{d}{dt}\Bigl( ln \rho({\bf q} ,{\bf p}, x)+
\beta(x) H({\bf q} ,{\bf p} ,x) \Bigr)=0, \]
i.e., the value $(ln \rho+\beta H)$ is a constant along of 
the trajectory of the system in 6N-dimensional phase space.
Let us denote this constant value by $\beta(x) {\cal F}(x)$.
Then we have
\[ ln \rho({\bf q} ,{\bf p} ,x)+
\beta(x) H({\bf q} ,{\bf p} ,x)=\beta (x){\cal F}(x),\]
where $d{\cal F}(x)/dt=0$. 
As a result, we get a canonical distribution function
\[ \rho({\bf q} ,{\bf p} ,x)=\exp \beta(x) \Bigl(
{\cal F}(x)- H({\bf q} ,{\bf p} ,x) \Bigr) . \]
The value ${\cal F}(x)$ is defined
by the normalization condition.
Therefore the distribution of this few-particle system is
a canonical distribution.

\subsection{Non-Canonical Distributions for Few-Particle Systems}

In Sec 3.3, we consider $\beta(H,x)=\beta(x)$. 
Let us consider the linear function $\beta=\beta(H,x)$. \\

{\bf Corollary 2.}
{\it The linear function $\beta(H,x)$ in the form
\[ \beta(H,x)=\beta_1(x)+\beta_2(x)H \]
leads to the following non-canonical distribution function}
\[ \rho({\bf q},{\bf p},x)=\frac{1}{Z(x)} 
\exp -\Bigl( \beta_1(x)H+\frac{1}{2}\beta_2(x)H^2 \Bigr) .\]

The proof of this proposition can be directly derived from 
Eqs. (\ref{solution}) and (\ref{L}).

The well known non-Gaussian distribution is the Breit-Wigner distribution.
This distribution has a probability density function in the form 
$\rho(x)=1 / \pi(1+x^2)$. 
The Breit-Wigner distribution is also known in statistics 
as Cauchy distribution.  
The Breit-Wigner distribution is a generalized form 
originally introduced \cite{Breit36} to describe 
the cross-section of resonant nuclear scattering in the form
\be \label{BWD}
\rho(H({\bf q},{\bf p},x))=\frac{\lambda}{(H({\bf q},{\bf p},x)-E)^2+(\Gamma/2)^2} .
\ee
This distribution can be
derived from the transition probability of a resonant 
state with known lifetime \cite{Bohr69,Fermi51,Paul69}. 
The second non-Gaussian distribution, which is considered in this section,  
is classical Fermi-Bose distribution that
was suggested by Ebeling in Refs. \cite{Eb,SET}.
This distribution has the form 
\be \label{FermiBose2}
\rho(H({\bf q},{\bf p},x))=
\frac{1}{\exp [\beta(x) (H({\bf q},{\bf p},x)-\mu)]+a} .
\ee

{\bf Corollary 3.}
{\it If the function $\beta(H,x)$ of the non-holonomic constrain
is defined by
\be \beta(H,x)=\frac{2(H-E)}{(H-E)^2+(\Gamma/2)^2} ,
\ee
then we have thermodynamic few-particle systems with the Breit-Wigner 
distribution (\ref{BWD}).} \\

{\bf Corollary 4.}
{\it If the function $\beta(H,x)$ of the non-holonomic constrain has the form
\be
\beta(H,x)=\frac{\beta(x)}{1+\alpha \, \exp \ \beta(x) H},
\ee
then we have thermodynamic few-particle systems with classical Fermi-Bose 
distribution (\ref{FermiBose2}). } \\

Note that Ebeling derives the Fermi-Bose distribution function as 
a solution of the Fokker-Planck equation. 
It is known that Fokker-Planck equation can be derived
from the Liouville equation \cite{Is}.
We derive classical Fermi-Bose distribution  
as a solution of the Liouville equation.

If the non-potential forces ${\bf F}^{(n)}_i$ are determined 
by the Hamiltonian 
\be {\bf F}^{(n)}_i=-\partial G(H)/ \partial {\bf p}_i , \ee
then we have the thermodynamic few-particle systems,
which are considered in Refs. \cite{Eb,SET}. These systems 
are called canonical dissipative systems.

Let us assume that Eq. (\ref{rhoH}) can be solved in the form 
\be H=T(x) G(\rho) , \ee
where $G$ depends on the distribution $\rho$. The function
$T(x)$ is a function of the parameters $x$.  
In this case, the function $\beta(H,x)$ is a composite function
\be
C(\rho)=-\beta(T(x)G(\rho),x). 
\ee
This function can be defined by
\be
C(\rho)=\frac{1}{\rho}\Bigl(T(x)
\frac{\partial G(\rho)}{\partial \rho}\Bigr)^{-1}.
\ee
As a result, we have the Liouville equation for the 
few-particle system in the form
\be
\frac{d\rho}{dt}=C(\rho) {\cal P} .
\ee
This equation is a nonlinear equation.
For example, the classical Fermi-Bose systems 
\cite{Eb} have the function in the form
\be
C(\rho)=-\beta(x)(\rho-s \rho^2) .
\ee
Note that the nonlinear evolution of statistical systems
is considered in Refs. \cite{nn1,nn2,nn3,nn4,nn5,nn6,nn7,nn8}.

\section{Non-Holonomic Constraint}

\subsection{Formulation of the Result}

In this section, 
we formulate the proposition, which allows us to derive 
the thermodynamic few-particle systems from any equations of 
motion of N-particle systems.
The aim of this section is to prove the following result. \\
 
{\bf Proposition 2.}
{\it For any few-particle system, which is defined by the equation
\be \frac{d{\bf q}_i}{dt}=\frac{\partial H}{\partial {\bf p}_i},\quad
\frac{d{\bf p}_i}{dt}=-\frac{\partial H}{\partial {\bf q}_i}
+{\bf F}^{(n)}_i, \quad i=1,...,N, \ee
there exists a thermodynamic few-particle system 
that is defined by the equations
\be \frac{d{\bf q}_i}{dt}=\frac{\partial H}{\partial {\bf p}_i},\quad
\frac{d{\bf p}_i}{dt}={\bf F}^{new}_i , \ee
and the distribution (\ref{distr}), 
where the non-potential forces ${\bf F}^{new}_i$ are defined by
\be \label{new2}
{\bf F}^{new}_i=
\frac{ {\bf P}_k{\bf P}_k \delta_{ij}-{\bf P}_i{\bf P}_j }{
{\bf P}_k{\bf P}_k} \Bigl(-\frac{\partial H}{\partial {\bf q}_j}
+{\bf F}^{(n)}_j \Bigr)
-\frac{{\bf P}_i {\bf Q}_j }{{\bf P}_k{\bf P}_k} 
\frac{\partial H}{\partial {\bf p}_j} . \ee
The vectors ${\bf P}_i$ and ${\bf Q}_i$ are defined by the equations
\[ {\bf P}_i=\frac{\partial \beta(H,x)}{\partial H} 
\frac{\partial H}{\partial {\bf p}_i} 
\frac{\partial H}{\partial {\bf p}_j}{\bf F}^{(n)}_j 
+\beta(H,x)\frac{\partial {\bf F}^{(n)}_j}{\partial {\bf p}_i} 
\frac{\partial H}{\partial {\bf p}_j} + \]
\be \label{51} +\beta(H,x){\bf F}^{(n)}_j\frac{\partial^2 H}{\partial 
{\bf p}_i\partial {\bf p}_j}
-\frac{\partial^2 {\bf F}^{(n)}_j}{\partial {\bf p}_i\partial {\bf p}_j} ,
\ee
and
\[ {\bf Q}_i=\frac{\partial \beta(H,x)}{\partial H}
\frac{\partial H}{\partial {\bf q}_i}
\frac{\partial H}{\partial {\bf p}_j}{\bf F}^{(n)}_j+ 
\beta(H,x)\frac{\partial {\bf F}^{(n)}_j}{\partial {\bf q}_i} 
\frac{\partial H}{\partial {\bf p}_j} + \]
\be \label{52}
+\beta(H,x){\bf F}^{(n)}_j\frac{\partial^2 H}{\partial 
{\bf q}_i\partial {\bf p}_j}
-\frac{\partial^2 {\bf F}^{(n)}_j}{\partial {\bf q}_i\partial {\bf p}_j}
 . \ee
}

Here we use the following notations
\[ {\bf a}_i {\bf b}_j {\bf c}_j={\bf a}_i\sum^N_{j=1} (b_{xj} c_{xj}+
b_{yj} c_{yj}+b_{zj} c_{zj}) . \] 

Note that the forces that are defined by Eqs. (\ref{new2}), 
(\ref{51}) and (\ref{52}) satisfy the non-holonomic constraint 
(\ref{NC-P0}), i.e.,
\be \label{NC-new} 
\frac{\partial {\bf F}^{new}_j}{\partial {\bf p}_j}
+\frac{\partial^2 H}{\partial {\bf q}_j \partial {\bf p}_j}
-\beta(H,x) {\bf F}^{new}_j \frac{\partial H}{\partial {\bf p}_j}
=0, \ee
where we use the omega function in the form (\ref{omega}).


\subsection{Proof of Proposition 2.}

Let us prove Eq. (\ref{new2}). 
Let us consider the N-particle classical system in 
the Hamilton picture.
Denote the position of the $i$th particle 
by ${\bf q}_i$ and its momentum by ${\bf p}_i$. 
Suppose that the system is subjected to a non-holonomic 
(non-integrable) constraint in the form 
\be \label{NC} 
f({\bf q},{\bf p},x)=0 . \ee
Differentiation of Eq. (\ref{NC}) with respect to time gives a relation
\be \label{TD}
{\bf P}_i({\bf q},{\bf p},x) \frac{d{\bf p}_i}{dt}+
{\bf Q}_i({\bf q},{\bf p},x)\frac{d{\bf q}_i}{dt}=0,
\ee
where the functions ${\bf P}_i$ and ${\bf Q}_i$ 
are defined by the equations
\be \label{AB} 
{\bf P}_i({\bf q},{\bf p},x)=\frac{\partial f}{\partial {\bf p}_i}, \quad
{\bf Q}_i({\bf q},{\bf p},x)=\frac{\partial f}{\partial {\bf q}_i}.
\ee
An unconstrained motion of the $i$th particle,
where $i=1,...,N$, is described by the equations
\be \label{EM1}\frac{d{\bf q}_i}{dt}={\bf K}_i , \quad
\frac{d{\bf p}_i}{dt}={\bf F}_i,\ee
where ${\bf F}_i$ is a resulting force, which acts on the $i$th particle.

The unconstrained motion gives a trajectory which leaves the constraint 
hypersurface (\ref{NC}).
The constraint forces ${\bf R}_i$ must be added to the equation 
of motion to prevent the deviation from the constraint hypersurface:
\be \label{EM2} \frac{d{\bf q}_i}{dt}={\bf K}_i , \quad
\frac{d{\bf p}_i}{dt}={\bf F}_i+{\bf R}_i .\ee
The constraint force ${\bf R}_i$ for the non-holonomic 
constraint is proportional to the ${\bf P}_i$ \cite{Dob}:
\be
{\bf R}_{i}=\lambda {\bf P}_i ,
\ee
where the coefficient $\lambda$ of the constraint force term 
is an undetermined Lagrangian multiplier.
For the non-holonomic constraint (\ref{NC}),  
the equations of motion (\ref{EM1}) are modified as
\be \label{EM3} \frac{d{\bf q}_i}{dt}={\bf K}_i , \quad
\frac{d{\bf p}_i}{dt}={\bf F}_i+\lambda{\bf P}_i .\ee

The Lagrangian coefficient $\lambda$ is determined 
by Eq. (\ref{TD}).
Substituting Eq. (\ref{EM2}) into Eq. (\ref{TD}), we get
\be \label{TD2}
{\bf P}_i ({\bf F}_i+\lambda{\bf P}_i)+
{\bf Q}_i {\bf K}_i=0 . \ee
Therefore the Lagrange multiplier $\lambda$ is equal to
\be \label{TD5}
\lambda =-\frac{{\bf P}_i {\bf F}_i
+{\bf Q}_i {\bf K}_i }{{\bf P}_k{\bf P}_k } .
\ee
As a result, we obtain the following equations
\be \label{EM4} 
\frac{d{\bf q}_i}{dt}={\bf K}_i, \quad
\frac{d{\bf p}_i}{dt}={\bf F}_i
-{\bf P}_i\frac{{\bf P}_j {\bf F}_j
+{\bf Q}_j  {\bf K}_j }{ {\bf P}_k{\bf P}_k}.\ee
These equations we can rewrite in the form (\ref{EM1}) 
\be \label{EM6} \frac{d{\bf q}_i}{dt}={\bf K}_i, \quad
\frac{d{\bf p}_i}{dt}={\bf F}^{new}_i \ee
with the new forces
\be \label{new}
{\bf F}^{new}_i=
\frac{ {\bf P}_k{\bf P}_k \delta_{ij}-{\bf P}_i{\bf P}_j }{
{\bf P}_k{\bf P}_k} {\bf F}_j
-\frac{{\bf P}_i {\bf Q}_j }{{\bf P}_k{\bf P}_k} {\bf K}_j . \ee
In general, the forces ${\bf F}^{new}_i$ are non-potentials forces 
(see examples in Ref. \cite{mplb}).

Equations (\ref{EM4}) are equations of 
the {\it holonomic} system. For any trajectory 
of the system in the phase space, we have $f=const$.
If initial values ${\bf q}_k(0)$ and ${\bf p}_k(0)$ satisfy 
the constraint condition $f({\bf q}(0),{\bf p}(0),x)=0$,
then solution of Eqs. (\ref{EM4}) and (\ref{new}) 
is a motion of the non-holonomic system.


Let us prove Eqs. (\ref{51}) and (\ref{52}).
In order to prove these equations, 
we consider the few-particle system (\ref{EM1}) with
\be
{\bf K}_i=\frac{\partial H}{\partial {\bf p}_i}, \quad
{\bf F}_i=-\frac{\partial H}{\partial {\bf q}_i}+{\bf F}^{(n)}_i, 
\ee
and the special form of the non-holonomic constraint (\ref{NC}).
Let us assume the following constraint: 
the velocity of the elementary phase volume change 
\ $\Omega({\bf q},{\bf p},x)$
is directly proportional to the power 
${\cal P}({\bf q},{\bf p},x)$
of the non-potential forces, i.e.,  
\be \label{NC2} 
\Omega({\bf q},{\bf p},x)=
\beta(H,x) {\cal P}({\bf q},{\bf p},x), \ee
where $\beta(H,x)$ depends on the Hamiltonian $H$.
Therefore the system is subjected to non-holonomic (non-integrable)
constraint (\ref{NC}) in the form 
\be \label{fPO} 
f({\bf q},{\bf p},x)= \beta(H,x) {\cal P}({\bf q},{\bf p},x)
-\Omega({\bf q},{\bf p},x)=0 . \ee
This constraint is a generalization of the condition 
which is suggested in Ref. \cite{mplb}. 
The power ${\cal P}$ of the 
non-potential forces ${\bf F}^{(n)}_i$ is defined by Eq. (\ref{power}). 
The function $\Omega$ is defined by Eq. (\ref{omega}).
As a result, we have Eq. (\ref{fPO}) for the non-potential forces in the form  
\[ \beta(H,x) {\bf F}^{(n)}_j \frac{\partial H}{\partial {\bf p}_j} -
\frac{\partial {\bf F}^{(n)}_j}{\partial {\bf p}_j}=0. \]
The functions ${\bf P}_i$ and ${\bf Q}_i$ for this constraint
can be find by differentiation of constraint. 
Differentiation of the function $f({\bf q},{\bf p},x)$ 
with respect to ${\bf p}_i$ gives
\[ {\bf P}_i({\bf q},{\bf p},x)=\frac{\partial f}{\partial {\bf p}_i} = 
\frac{\partial}{\partial {\bf p}_i} \Bigl(
\beta(H,x) {\bf F}^{(n)}_j \frac{\partial H}{\partial {\bf p}_j}\Bigr) 
-\frac{\partial}{\partial {\bf p}_i}
\frac{\partial {\bf F}^{(n)}_j}{\partial {\bf p}_j}  . \]
This expression leads us to Eq. (\ref{51}).
Differentiation of the function $f({\bf q} ,{\bf p} ,t)$ 
with respect to ${\bf q}_i$ gives
\[ {\bf Q}_i({\bf q},{\bf p},x)=\frac{\partial f}{\partial {\bf q}_i} = 
\frac{\partial}{\partial {\bf q}_i} \Bigl(
\beta(H,x) {\bf F}^{(n)}_j \frac{\partial H}{\partial {\bf p}_j}\Bigr) 
-\frac{\partial}{\partial {\bf q}_i}
\frac{\partial {\bf F}^{(n)}_j}{\partial {\bf p}_j} . \]
This expression leads to Eq. (\ref{52}).

\subsection{Few-Particle Systems with Minimal Constraint}

Let us consider the simple constraints for
to realize the classical systems with 
canonical and non-canonical distributions. 
Let us consider few-particle system, which
is defined by the simplest form of the Hamiltonian 
\be \label{S1}
H({\bf q},{\bf p},x)=\frac{{\bf p}^2}{2m}+U({\bf q},x) ,
\ee
and the non-potential forces
\be \label{S2}
{\bf F}^{(n)}_i=-\gamma {\bf p}_i .
\ee
Here ${\bf p}^2=\sum^N_{i=1}{\bf p}^2_i$, 
and $N$ is a number of particles. 
For the minimal constraint models, the non-holonomic constraint 
is defined by the equation
\be
f({\bf q},{\bf p},x)=\beta(H,x)\frac{{\bf p}^2}{m}-3N=0 .
\ee
The phase space gradients (\ref{51}) and (\ref{52}) 
of this constraint are represented in the form
\[ {\bf P}_i=\left(\frac{\partial \beta(H,x)}{\partial H}
\frac{{\bf p}^2}{2m}+\beta(H,x)\right) \frac{2{\bf p}_i}{m} ,
\quad
{\bf Q}_i=\frac{\partial \beta(H,x)}{\partial H}
\frac{\partial H}{\partial {\bf q}_i} . \]
The non-potential forces of the minimal constraint models
have the form
\[ {\bf F}^{new}_i=-
\frac{{\bf p}^2\delta_{ij}-{\bf p}_i{\bf p}_j}{{\bf p}^2}
\frac{\partial U}{\partial {\bf q}_j}
+\frac{{\bf p}_i{\bf p}_j }{2{\bf p}^2
([{\bf p}^2/2m) \partial \beta(H,x)/ \partial H]+\beta(H,x))} 
\frac{\partial \beta(H,x)}{\partial H} \frac{\partial U}{\partial {\bf q}_j} . \]
It is easy to see that all minimal constraint models 
have the potential forces. 

Note that the minimal Gaussian constraint model is characterized by
\[ \frac{\partial \beta(H,x)}{\partial H}=0, \]
In this case, we have the non-potential forces in the form
\[ {\bf F}^{new}_i=-
\frac{{\bf p}^2\delta_{ij}-{\bf p}_i{\bf p}_j}{{\bf p}^2}
\frac{\partial U}{\partial {\bf q}_j} . \]
The few-particle systems are the constant temperature systems
that are considered in Refs. 
\cite{E,EHFML,HG,EM,Nose1,Nose2,Nose,Tuck2,mplb}. 

\subsection{Few-Particle Systems with Minimal Gaussian Constraint}

Let us consider the N-particle system with the Hamiltonian (\ref{S1}), 
the function $\beta(H,x)=3N/kT(x)$, 
and the linear friction force (\ref{S2}).
Substituting Eq. (\ref{S2}) into Eqs. (\ref{power}) 
and (\ref{omega}), we get
the power ${\cal P}$ and the omega function $\Omega$:
\[ {\cal P}=-\frac{\gamma}{m} {\bf p}^2, \quad \Omega=-3\gamma N. \]
In this case, the non-holonomic constraint has the form
\be \label{PO2} \frac{{\bf p}^{2}}{m}=kT(x), \ee
i.e., the kinetic energy of the system must be a constant.
Note that Eq. (\ref{PO2}) has not the friction parameter $\gamma$. 

For the few-particle system with friction (\ref{S2})
and non-holonomic constraint (\ref{PO2}), we have 
the following equations of motion 
\be \label{em} 
\frac{d{\bf q}_i}{dt}= \frac{{\bf p}_i}{m} , \quad
\frac{d{\bf p}_i}{dt}=-\frac{\partial U}{\partial {\bf q}_i}
-\gamma {\bf p}_i+
\lambda \frac{\partial f}{\partial {\bf p}_i}, \ee
where the function $f=f({\bf q},{\bf p})$ is defined by
\be \label{con} f({\bf q},{\bf p})=
\frac{1}{2}\Bigl({\bf p}^{2}-mkT(x) \Bigr):
\quad f({\bf q},{\bf p})=0. \ee
Equation (\ref{em}) and condition (\ref{con})
define 6N+1 variables $({\bf q},{\bf p},\lambda)$.
Let us find the Lagrange multiplier $\lambda$.
Substituting Eq. (\ref{con}) into Eq. (\ref{em}), we get
\be \label{em2}
\frac{d{\bf p}_i}{dt}=-\frac{\partial U}{\partial {\bf q}_i}
+(\lambda-\gamma) {\bf p}_i . \ee
Using $df/dt=0$ in the form 
\be \label{pp0} {\bf p}_i \frac{d{\bf p}_i}{dt}=0 \ee
and substituting  Eq. (\ref{em2}) into 
Eq. (\ref{pp0}), we get
the Lagrange multiplier $\lambda$ in the form
\[ \lambda= \frac{1}{mkT(x)}
{\bf p}_j\frac{\partial U}{\partial {\bf q}_j}+\gamma . \]
As a result, we have the holonomic system 
that is defined by the equations
\be \label{em4} \frac{d{\bf q}_i}{dt}=\frac{{\bf p}_i}{m} , \quad
\frac{d{\bf p}_i}{dt}= \frac{1}{mkT(x)}
{\bf p}_i {\bf p}_j \frac{\partial U}{\partial {\bf q}_j}
-\frac{\partial U}{\partial {\bf q}_i}. \ee
For the few-particle system (\ref{em4}), 
condition (\ref{PO2}) is satisfied.
If the time evolution of the few-particle system 
is defined by Eq. (\ref{em4}),
then we have the canonical distribution function in the form
\be \label{cdf} 
\rho({\bf q},{\bf p},x)=
\frac{1}{Z(x)} \exp -\frac{H({\bf q},{\bf p},x)}{kT(x)} . \ee
where $Z(x)$ is defined by the normalization condition. 
For example, the few-particle system with the forces
\be {\bf F}_i=-m\omega^2(x) {\bf q}_i+ \frac{\omega^2(x)}{kT(x)} {\bf p_i}
({\bf p}_j{\bf q}_j) \ee
has canonical distribution (\ref{cdf}) of 
the linear harmonic oscillator with Hamiltonian
\[ H({\bf q},{\bf p},x)=
\frac{{\bf p}^2}{2m}+\frac{m\omega^2(x) {\bf q}^2}{2}. \]

\section{Conclusion}

In this paper we derive the extension of the statistical
thermodynamics to the wide class of few-particle systems.
We consider few-particle systems with distributions 
that are defined by Hamiltonian and Liouville equation.
These systems are described by the non-holonomic (non-integrable)
constraint \cite{mplb}: the velocity of the elementary phase volume change 
is directly proportional to the power of non-potential forces.  
In the general case, the coefficient of this proportionality 
is defined by Hamiltonian. 
This constraint allows us to derive the distribution 
function of the few-particle system, even in far-from equilibrium states.
The few-particle systems that have some analog of the thermodynamic
laws is characterized by the distribution functions
that are determined by the Hamiltonian. 
The examples of these few-particle systems are 
the constant temperature systems \cite{E,EHFML,HG,EM,Nose1,Nose2,Nose,Tuck2},
the canonical-dissipative systems \cite{Eb,SET}, 
and the Fermi-Bose classical systems \cite{Eb}.
For the few-particle systems, we can use
the analogs of the usual thermodynamics laws.  
Note that the number of particles  is an arbitrary natural number since 
we do not use the condition $N\gg 1$ or $N \rightarrow \infty$.
This allows one to use the suggested few-particle systems 
for the simulation \cite{FS} for the molecular dynamics.

The quantization of the evolution equations 
for non-Hamiltonian and dissipative 
systems was suggested in Refs. \cite{Tarpla1,Tarmsu}.
Using this quantization it is easy to derive the 
quantum analog of few-particle systems  
that leads to some analog of the thermodynamic laws.
We can derive the canonical and non-canonical statistical operators \cite{Tarkn1}
that are determined by the Hamiltonian \cite{Tarpre02,Tarpla02}.
The suggested few-particle systems 
can be generalized by the quantization 
method that is considered in Refs. \cite{Tarpla1,Tarmsu}.



\end{document}